\documentclass[12pt]{article}

\begin{document}
\begin{titlepage}
\begin{flushright}
UFIFT-QG-04-2 \\ gr-qc/0408002
\end{flushright}
\vspace{.4cm}
\begin{center}
\textbf{De Sitter Breaking in Field Theory}
\end{center}
\begin{center}
R. P. Woodard$^{\dagger}$
\end{center}
\begin{center}
\textit{Department of Physics \\ University of Florida \\
Gainesville, FL 32611 USA}
\end{center}

\begin{center}
ABSTRACT
\end{center}
I argue against the widespread notion that manifest
de Sitter invariance on the full de Sitter manifold is either useful 
or even attainable in gauge theories. Green's functions and propagators 
computed in a de Sitter invariant gauge are generally more complicated 
than in some noninvariant gauges. What is worse, solving the gauge-fixed 
field equations in a de Sitter invariant gauge generally leads to 
violations of the original, gauge invariant field equations. The most
interesting free quantum field theories possess no normalizable, de 
Sitter invariant states. This precludes the existence of de Sitter 
invariant propagators. Even had such propagators existed, infrared 
divergent processes would still break de Sitter invariance.

\begin{flushleft}
\end{flushleft}
\vspace{.4cm}
\begin{flushleft}
$^{\dagger}$ e-mail: woodard@phys.ufl.edu
\end{flushleft}
\end{titlepage}

\section{Introduction}

Stanley Deser has been my mentor for over two decades. One of the many
things he taught me is that theoretical physics is characterized by long
periods of stagnation, punctuated by bursts of activity after some insight 
or technical advance makes progress possible. When this happens one has to 
push forward as far and as fast as possible because these opportunities 
don't arise often. Stanley's career has exemplified this, starting in the 
late 50's with the canonical formulation of gravity that his work with 
Arnowitt and Misner made possible 
\cite{AD,ADM1,ADM2,ADM3,ADM4,ADM5,ADM6,ADM7,ADM8,ADM9,ADM10,ADM10,ADM11,ADM12}.
Another fine example is the way Stanley and various collaborators exploited 
the newly developed technology of dimensional regularization and the 
background field formalism in the mid 70's to analyze the one loop divergences 
of gravity combined with other theories \cite{DN1,DN2,DTN1,DTN2}.

Much of my recent work has dealt with exploiting a technical advance that
has made it possible to get interesting results from quantum field theory
during inflation. The advance is the development of relatively simple 
propagators for massless fields on a locally de Sitter background of 
arbitrary spacetime dimension. This has made it possible to use dimensional
regularization to go beyond the coincidence limits of one loop stress tensors 
--- the technology for which had been codified before I graduated \cite{BD}. 
One can now get at the deeply nonlocal, ultraviolet finite parts of quantum 
processes during inflation.

Section 2 of this article reviews what has been done. Section 3 explains
why some of my methods offend the aesthetic prejudices of the mathematically 
minded. However much more attractive the formalism might seem their way, it 
would be neither practical, nor physically correct, nor would its most 
interesting predictions be free of the unaesthetic properties of my 
techniques. The various problems of practicality and of principle are 
described in section 4. Section 5 summarizes my conclusions.

\section{Quantum Field Theory during Inflation}

I model inflation using a portion of the full de Sitter manifold known as 
the open conformal coordinate patch. If the $D$-dimensional cosmological 
constant is $\Lambda \equiv (D\!-\!1) H^2$, the invariant element is,
\begin{equation}
ds^2 = a^2 \Bigl(-d\eta^2 + d\vec{x} \cdot d\vec{x}\Bigr) \qquad {\rm where}
\qquad a(\eta) = -\frac1{H \eta} \; . \label{ds^2}
\end{equation}
The conformal time $\eta$ runs from $-\infty$ to zero. The various propagators
have simple expressions in terms of the following function of the invariant
length $\ell(x;x')$ between $x^{\mu}$ and $x^{\prime \mu}$,
\begin{equation}
y(x;x') \equiv 4 \sin^2\Bigl(\frac12 H \ell(x;x')\Bigr) = a a' H^2 
\Bigl(\Vert \vec{x} \!-\! \vec{x}' \Vert^2 \!-\! (\vert\eta \!-\! \eta'
\vert - i \delta)^2\Bigr) \; .
\end{equation}

One might expect that the inflationary expansion of this spacetime makes
quantum effects stronger by allowing virtual particles to persist longer
than in flat space. Indeed, it is simple to see that any sufficiently
long wavelength ($\lambda > 1/H$) virtual particle which is massless 
on the Hubble scale can exist forever \cite{RPW}. However, one must also 
consider the {\it rate} at which virtual particles emerge from the vacuum. 
Classical conformal invariance causes this rate to fall off exponentially, 
so any long wave length virtual particles which emerge become real, but 
very few emerge \cite{RPW}. To get enhanced quantum effects during 
inflation requires quanta which are effectively massless and also not 
conformally invariant. Even one such particle can catalyze processes
involving conformally invariant particles.

It has long been known how to write the propagator for a massless,
conformally coupled scalar in arbitrary dimension \cite{BD},
\begin{equation}
{i\Delta}_{\rm cf}(x;x') = \frac{H^{D-2}}{(4\pi)^{\frac{D}2}} \Gamma\Bigl(
\frac{D}2 \!-\! 1\Bigr) \Bigl(\frac4{y}\Bigr)^{\frac{D}2-1} \; .
\end{equation}
Massless fermions are also conformally invariant in any dimension and
their propagator is closely related,
\begin{equation}
i\Bigl[{}_iS_j\Bigr](x;x') = (a a')^{\frac{1-D}2} i \partial_{\mu} \gamma^{
\mu}_{ij} \Bigl[ (a a')^{\frac{D}2 - 1} i \Delta_{\rm cf}(x;x')\Bigr] \; .
\end{equation}
One can compute with these propagators but the results are not much different 
from flat space on account of conformal invariance. 

The advance that has made interesting quantum effects computable is explicit 
expressions for the propagators of particles which are massless and not 
conformally invariant. The first of these is the minimally coupled scalar,
\vfill\eject
\begin{eqnarray}
\lefteqn{i \Delta_A(x;x') =  i \Delta_{\rm cf}(x;x') } \nonumber \\
& & + \frac{H^{D-2}}{(4\pi)^{\frac{D}2}} \frac{\Gamma(D \!-\! 1)}{\Gamma(
\frac{D}2)} \left\{\! \frac{D}{D\!-\! 4} \frac{\Gamma^2(\frac{D}2)}{\Gamma(D
\!-\! 1)} \Bigl(\frac4{y}\Bigr)^{\frac{D}2 -2} \!\!\!\!\!\! - \pi 
\cot\Bigl(\frac{\pi}2 D\Bigr) + \ln(a a') \!\right\} \nonumber \\
& & + \frac{H^{D-2}}{(4\pi)^{\frac{D}2}} \! \sum_{n=1}^{\infty}\! \left\{\!
\frac1{n} \frac{\Gamma(n \!+\! D \!-\! 1)}{\Gamma(n \!+\! \frac{D}2)} 
\Bigl(\frac{y}4 \Bigr)^n \!\!\!\! - \frac1{n \!-\! \frac{D}2 \!+\! 2} 
\frac{\Gamma(n \!+\!  \frac{D}2 \!+\! 1)}{\Gamma(n \!+\! 2)} \Bigl(\frac{y}4
\Bigr)^{n - \frac{D}2 +2} \!\right\} \! . \quad \label{DeltaA}
\end{eqnarray}
This might seem a daunting expression but it isn't so bad because the
infinite sum on the final line vanishes in $D=4$, and each term in the
series goes like positive powers of $y(x;x')$. This means the infinite
sum can only contribute when multiplied by a divergent term, and even
then only a small number of terms can contribute.

Fascinating physics has been revealed by endowing such a scalar 
with different sorts of interactions. When a quartic self-interaction
is present one can compute the VEV of the stress tensor \cite{OW1,OW2}
and the scalar self-mass-squared \cite{BOW} at one and two loop orders. 
The resulting model shows a violation of the weak energy condition --- on 
cosmological scales! --- in which inflationary particle production drives 
the scalar up its potential and induces a curious sort of time-dependent 
mass. When a complex scalar of this type is coupled to electromagnetism is 
has been possible to compute the one loop vacuum polarization \cite{PTW1,PTW2} 
and use the result to solve the quantum corrected Maxwell equations \cite{PW1}. 
Although photon creation is suppressed during inflation, this model shows 
a vast enhancement of the 0-point energy of super-horizon photons which may 
serve to seed cosmological magnetic fields \cite{DDPT,DPTD,PW2}. Finally, 
when a real scalar of this type is Yukawa coupled to a massless Dirac fermion 
it has been possible to compute the one loop fermion self-energy and use it 
to solve the quantum corrected Dirac equation \cite{PW3}. The resulting 
model shows explosive creation of fermions which should make inflation
end with the super-horizon modes in a degenerate Fermi gas!

Electromagnetism is a special case, being conformally invariant in $D=4$
but not generally. My favorite gauge fixing term is an analogue of the
one introduced by Feynamn in flat space,
\begin{equation}
\mathcal{L}_{GF} = -\frac12 a^{D-4} \Bigl(\eta^{\mu\nu} A_{\mu , \nu}
- (D \!-\!4) H a A_0\Bigr)^2 \; . \label{EM}
\end{equation}
Because space and time components are treated differently it is useful to 
have an expression for the purely spatial part of the Minkowski metric,
\begin{equation}
\overline{\eta}_{\mu\nu} \equiv \eta_{\mu\nu} + \delta^0_{\mu} \delta^0_{\nu}
\; .
\end{equation}
In this gauge the photon propagator takes the form,
\begin{equation}
i\Bigl[{}_{\mu} \Delta_{\nu}\Bigr](x;x') = \overline{\eta}_{\mu\nu} a a'
i\Delta_B(x;x') - \delta^0_{\mu} \delta^0_{\nu} a a' i\Delta_C(x;x') \; .
\end{equation}
The B-type and $C$-type propagators are,
\begin{eqnarray}
\lefteqn{i \Delta_B(x;x') =  i \Delta_{\rm cf}(x;x') - \frac{H^{D-2}}{(4
\pi)^{\frac{D}2}} \! \sum_{n=0}^{\infty}\! \left\{\!  \frac{\Gamma(n \!+\! D 
\!-\! 2)}{\Gamma(n \!+\! \frac{D}2)} \Bigl(\frac{y}4 \Bigr)^n \right. } 
\nonumber \\
& & \hspace{6.5cm} \left. - \frac{\Gamma(n \!+\!  \frac{D}2)}{\Gamma(n \!+\! 
2)} \Bigl( \frac{y}4 \Bigr)^{n - \frac{D}2 +2} \!\right\} \! , \qquad 
\label{DeltaB} \\
\lefteqn{i \Delta_C(x;x') =  i \Delta_{\rm cf}(x;x') + 
\frac{H^{D-2}}{(4\pi)^{\frac{D}2}} \! \sum_{n=0}^{\infty} \left\{\!
(n\!+\!1) \frac{\Gamma(n \!+\! D \!-\! 3)}{\Gamma(n \!+\! \frac{D}2)} 
\Bigl(\frac{y}4 \Bigr)^n \right. } \nonumber \\
& & \hspace{4.5cm} \left. - \Bigl(n \!-\! \frac{D}2 \!+\!  3\Bigr) \frac{
\Gamma(n \!+\! \frac{D}2 \!-\! 1)}{\Gamma(n \!+\! 2)} \Bigl(\frac{y}4 
\Bigr)^{n - \frac{D}2 +2} \!\right\} \! . \qquad \label{DeltaC}
\end{eqnarray}
As with the $A$-type propagator (\ref{DeltaA}), the infinite sums in
(\ref{DeltaB}) and (\ref{DeltaC}) vanish in $D=4$. In fact the $B$-type
and $C$ type propagators agree in $D=4$, and the photon propagator is the
same for $D=4$ as it is in flat space!

No results have been published using the photon propagator but L. D. Duffy 
and I are computing the one loop scalar self-mass-squared in
scalar QED. We expect its secular growth to eventually choke off the 
inflationary particle production that so enhances the one loop vacuum 
polarization \cite{PW1}. Of course this can't eliminate scalars which have
already been ripped out of the vacuum, or the vacuum polarization they
induce. A similar computation of the scalar self-mass-squared of the Yukawa 
scalar fails to show any secular growth at one loop order \cite{DW},
implying that the scalar-catalyzed production of super-horizon fermions goes
to completion \cite{PW3}.

Gravitons are also massless without conformal invariance. I define the
graviton field $\psi_{\mu\nu}(x)$ as follows,
\begin{equation}
g_{\mu\nu}(x) \equiv a^2 \Bigl(\eta_{\mu\nu} + \kappa \psi_{\mu\nu}(x)\Bigr)
\qquad {\rm where} \qquad \kappa^2 \equiv 16 \pi G \; .
\end{equation}
My favorite gauge fixing term is an analogue of the de Donder term used in 
flat space \cite{TW1},
\begin{equation}
\mathcal{L}_{GF} = -\frac12 a^{D-2} \eta^{\mu\nu} F_{\mu} F_{\nu} \; , \;
F_{\mu} \equiv \eta^{\rho\sigma} \Bigl(\psi_{\mu\rho , \sigma} 
- \frac12 \psi_{\rho \sigma , \mu} + (D \!-\! 2) H a \psi_{\mu \rho}
\delta^0_{\sigma} \Bigr) . \label{GR}
\end{equation}
With these definitions the graviton propagator takes the form of a sum
of three constant index factors times the three scalar propagators,
\begin{equation}
i\Bigl[{}_{\mu\nu} \Delta_{\rho\sigma}\Bigr](x;x') = \sum_{I=A,B,C}
\Bigl[{}_{\mu\nu} T^I_{\rho\sigma}\Bigr] i\Delta_I(x;x') \; . \label{gprop}
\end{equation}
The index factors are,
\begin{eqnarray}
\Bigl[{}_{\mu\nu} T^A_{\rho\sigma}\Bigr] & = & 2 \, \overline{\eta}_{\mu (\rho}
\overline{\eta}_{\sigma) \nu} - \frac2{D\!-\! 3} \overline{\eta}_{\mu\nu}
\overline{\eta}_{\rho \sigma} \; , \\
\Bigl[{}_{\mu\nu} T^B_{\rho\sigma}\Bigr] & = & -4 \delta^0_{(\mu} 
\overline{\eta}_{\nu) (\rho} \delta^0_{\sigma)} \; , \\
\Bigl[{}_{\mu\nu} T^C_{\rho\sigma}\Bigr] & = & \frac2{(D \!-\!2) (D \!-\!3)}
\Bigl[(D \!-\!3) \delta^0_{\mu} \delta^0_{\nu} + \overline{\eta}_{\mu\nu}\Bigr]
\Bigl[(D \!-\!3) \delta^0_{\rho} \delta^0_{\sigma} + \overline{\eta}_{\rho
\sigma}\Bigr] \; .
\end{eqnarray}

The full power of the dimensionally regulated graviton propagator has not so 
far been exploited in published work. However, the one loop graviton 
self-energy \cite{TW2} has been computed using a $D=4$ cutoff. The expectation 
value of the invariant element has also been obtained at two loop order 
\cite{TW3}. These results indicate that the back-reaction from graviton 
production slows inflation by an amount which eventually becomes 
nonperturbatively large \cite{TW4}. N. C. Tsamis and I have used the 
dimensionally regulated formalism to compute the expectation value of the
metric at one loop order. E. O. Kahya and I are also using it to compute 
the one loop scalar self-mass-squared induced by graviton exchange. This
might have important consequences for models which inflate for a very 
large number of e-foldings.

\section{What Bothers People}

Despite all the results that have been obtained, and the ones which are
attainable, the response of the theoretical physics community has been 
--- {\it underwhelming}. Different segments of the community have different 
reasons for ignoring my work. Many inflationary cosmologists feel that 
causality precludes interesting quantum field theoretic effects. Some of them 
even seem to have forgotten that the density perturbations which figure so 
prominently in recent observation \cite{WMAP1,WMAP2} are driven by 
precisely the same inflationary particle production \cite{AAS,MC} that 
underlies each of the effects reported in the previous section! String 
theorists are not much interested in physics that doesn't make essential 
use of their candidate for a theory of everything. They also flirt with 
the notion that there are no observables in de Sitter, which requires 
them to disbelieve that quantum corrections to the field equations mean 
anything. Loop space gravity people have trouble achieving correspondence 
with most forms of perturbation theory, including mine. And phenomenologists 
seek to work out the consequences of {\it popular} theories, so the fact 
that few people pay attention to my work serves to justify continuing to 
ignore it!

There isn't much I can do about this. But I {\it could} converse with one
segment of the community if only it was possible to overcome the distaste 
its members have for the methods I use. I refer to the mathematical 
relativists. They are prepared to accept that quantum field theory might
have interesting effects during inflation, and that these can be quantified 
in a reliable way. They are even willing to let me use Minkowski-signature 
perturbation theory starting with a prepared initial state! However, they 
are strongly attracted by the analogy between Minkowski space and de Sitter 
space, the maximally symmetric solutions of Einstein's equations for 
$\Lambda = 0$ and $\Lambda > 0$, respectively. They feel that manifest de 
Sitter invariance on the full de Sitter manifold should be as powerful an 
organizing principle for quantum field theory with $\Lambda > 0$ as 
Poincar\'e invariance has been for $\Lambda = 0$. So it bothers them that 
my open conformal coordinate patch (\ref{ds^2}) does not cover the full de 
Sitter manifold and that the gauge fixing terms I use --- (\ref{EM}) and 
(\ref{GR}) --- are not de Sitter invariant.

\section{You Can't Always Get What You Want}

Mick Jagger and Keith Richards are not my favorite authorities on much of 
anything, but one of their songs seems relevant here. I will argue that it 
isn't necessary, convenient or even possible to impose de Sitter invariant 
gauges and work on the full manifold. Nor would doing so lead to de Sitter 
invariant results for the most interesting processes if it were possible.

Necessity is the simplest issue. Everyone understands that it isn't
{\it necessary} to use de Sitter invariant gauges, just as it isn't
necessary to use Poincar\'e invariant gauges in flat space. Nor is 
there any logical problem with restricting physics to the open conformal 
coordinate patch (\ref{ds^2}), especially if one contemplates releasing
a prepared state from a finite initial time. The condition $\eta = {\rm
constant}$ defines a perfectly good Cauchy surface. Information from the
rest of the full de Sitter manifold can only propagate to the future of
such a surface by passing through it as part of the initial condition.
Indeed, the case for restricting to (\ref{ds^2}) can be put much more
strongly if one imagines --- as I do --- the local de Sitter background 
as merely a model for the more complicated geometry of the inflating 
epoch of cosmology. I am not interested in quantum field theory on perfect 
de Sitter space but rather in potentially observable quantum phenomena 
from the epoch of primordial inflation. In that case the relevant symmetries 
are homogeneity and isotropy, not the full de Sitter group, and the 
conformal coordinate patch --- with arbitrary $a(\eta)$ --- is the 
coordinate system in which these symmetries are manifest.

The reason people typically prefer to maintain manifest Poincar\'e 
invariance in flat space is that it makes things {\it simpler}. That
this is not true for de Sitter can be seen by comparing propagators in 
my gauges with those in the simplest de Sitter invariant gauges. It will 
sharpen the distinction if we take $D=4$. In that case the photon propagator 
in my gauge (\ref{EM}) is the same function of conformal coordinates as it 
is in flat space,
\begin{equation} 
i\Bigl[{}_{\mu} \Delta_{\nu}\Bigr](x;x')\Bigl\vert_{D=4}  = 
\frac{\eta_{\mu\nu}}{4 \pi^2 \Delta x^2} \; ,
\end{equation}
where ${\Delta x}^2 \equiv \Vert \vec{x} \!-\! \vec{x}' \Vert^2 \!-\! 
(\vert\eta \!-\! \eta' \vert - i \delta)^2$. It is worth pointing out 
that this expression applies to {\it any} homogeneous and isotropic
geometry in conformal coordinates, not just the special case of de
Sitter.

The simplest de Sitter invariant photon propagator of which I 
know was obtained by Allen and Jacobson \cite{AJ} with the gauge fixing 
term,
\begin{equation}
\mathcal{L}_{\rm inv} = -\frac12 \Bigl( g^{\mu\nu} A_{\mu ; \nu} \Bigr)^2 
\sqrt{-g} = -\frac12 a^{D-4} \Bigl(\eta^{\mu\nu} A_{\mu , \nu} - (D-2) H a 
A_0\Bigr)^2 \; .
\end{equation}
Their propagator takes the form,
\begin{equation}
i\Bigl[{}_{\mu} \Delta_{\nu}\Bigr](x;x')\Bigl\vert_{\rm inv} = \alpha(y)
\Bigl[{}_{\mu} g_{\nu}\Bigr](x;x') + \beta(y) \Bigl[{}_{\mu} n\Bigr](x;x') 
\Bigl[n_{\nu}\Bigr](x;x') \; , \label{AJprop}
\end{equation}
where $y(x;x') \equiv a a' H^2 {\Delta x}^2$, $\Bigl[{}_{\mu} g_{\nu}
\Bigr](x;x')$ is the parallel transport matrix and $\Bigl[{}_{\mu} n
\Bigr](x;x')$ and $\Bigl[n_{\nu} \Bigr](x;x')$ are the gradients with respect 
to $x^{\mu}$ and $x^{\prime \nu}$ of the geodesic length. In $D=4$ the 
coefficient functions are,
\begin{eqnarray}
\alpha(y) & = & \frac{H^2}{4 \pi^2} \left\{ \frac1{y} + \frac{\frac13}{4\!-\!y}
+ \frac{(4 \!-\! \frac23 y)}{(4 \!-\! y)^2} \ln\Bigl(\frac{y}4\Bigr) \right\}
\; , \\
\beta(y) & = & \frac{H^2}{4 \pi^2} \left\{- \frac{\frac16 y}{4\!-\!y}
- \frac{\frac23 y}{(4 \!-\! y)^2} \ln\Bigl(\frac{y}4\Bigr) \right\} \; .
\end{eqnarray}
The $3\!+\! 1$ decomposition of the parallel transport matrix is,
\begin{equation}
\Bigl[{}_{\mu} g_{\nu}\Bigr] = a a' \left( 
\matrix{1 & 0 \cr 0 & \delta_{mn}} \right) + 
\frac2{4 \!-\! y} \left( \matrix{ -(a \!+\! a')^2 &
(a \!+\! a') a a' H {\Delta x}_n \cr -(a \!+\! a') a a' H {\Delta x}_m &
a^2 a^{\prime 2} H^2 {\Delta x}_m {\Delta x}_n} \right) , \qquad
\end{equation}
The other tensor has the following $3 \!+\! 1$ decomposition,
\begin{eqnarray}
\lefteqn{\Bigl[{}_{\mu} n\Bigr] \Bigl[n_{\nu}\Bigr] = -\frac1{y} \left( 
\matrix{a a' y \!+\! 2a^2 \!+\! 2 a^{\prime 2} & -2 a^2 a' H {\Delta x}_n
\cr 2 a a^{\prime 2} H {\Delta x}_m & 0} \right) } \nonumber \\
& & \hspace{2cm} + \frac4{y (4 \!-\! y)}
\left( \matrix{ (a \!+\! a')^2 & -(a \!+\! a') a a' H {\Delta x}_n \cr 
(a \!+\! a') a a' H {\Delta x}_m & a^2 a^{\prime 2} H^2 {\Delta x}_m 
{\Delta x}_n } \right) \; .
\end{eqnarray}
However much one may admire manifest de Sitter invariance, I hope we can 
all agree that it doesn't simplify propagators.

But suppose you are fanatical about de Sitter invariance and you prefer to
compute on the full de Sitter manifold in a gauge which is manifestly de 
Sitter invariant, no matter how much harder it is. {\bf In that case you 
risk violating the invariant equations of motion!} The problem arises from 
combining the causal properties of de Sitter with the constraint equations
of any gauge theory. Before gauge fixing the constraint equations are 
elliptic, and they typically result in a nonzero response to sources
throughout the de Sitter manifold, even in regions which are not 
future-related to the source. But gauge fixing in a de Sitter invariant 
manner results in hyperbolic equations for which the response to sources 
is zero for regions which are not future-related to the source. As far
as I know this problem was first noted by Penrose \cite{P} in 1963. Tsamis 
and I encountered it for gravity in 1994 \cite{TW1} and recent studies for 
electromagnetism have been conducted by Bi\v{c}\'ak and Krtou\v{s} \cite{BK}.

To better understand the problem let us adopt the standard closed 
coordinatization of the full de Sitter manifold,
\begin{equation}
ds^2 = -dt^2 + H^{-2} \cosh^2(Ht) \Bigl( d\chi^2 + \sin^2(\chi) d\theta^2
+ \sin^2(\chi) \sin^2(\theta) d\phi^2\Bigr) \; .
\end{equation}
Consider the invariant Maxwell equations for a pair of oppositely
charged point particles,
\begin{equation}
\partial_{\mu} \Bigl(\sqrt{-g} g^{\mu\rho} g^{\nu\sigma} F_{\rho \sigma}\Bigr)
= q \int d\tau \Bigl[ \dot{z}^{\nu}_+(\tau) \delta^4\Bigl(x \!-\! z_+(\tau)
\Bigr) - \dot{z}^{\nu}_-(\tau) \delta^4\Bigl(x \!-\! z_-(\tau) \Bigr)\Bigr] .
\end{equation}
When the $+q$ charge is stationary at $\chi = 0$ and the $-q$ charge is 
stationary at $\chi = \pi$ a perfectly good solution exists,
\begin{equation}
A_{\mu} = \delta^0_{\mu} A_0(t,\chi) = \frac{q H}{4\pi} {\rm sech}(Ht) 
\cot(\chi) \; . \label{soln}
\end{equation}
Suppose we try to find a gauge parameter $\theta(x)$ such that the
transformed field, $A^{\prime}_{\mu} = A_{\mu} - \partial_{\mu} \theta$
obeys the de Sitter invariant condition $A^{\prime \mu}_{~~;\mu} = 0$,
\begin{equation}
\partial_{\mu} \! \Bigl(\! \sqrt{-g} g^{\mu\nu} \partial_{\nu} \theta \!\Bigr) 
\!=\! \partial_{\mu} \! \Bigl(\! \sqrt{-g} g^{\mu\nu} A_{\nu} \!\Bigr) \!=\!
\frac{-q}{4\pi H} \sinh(2Ht) \frac{\sin^3(\chi)}{\cos(\chi)} \sin(\theta) 
\! \equiv \! S(x) . \label{source}
\end{equation}
You might think this is easy with a Green's function,
\begin{equation}
\partial_{\mu} \Bigl( \sqrt{-g} g^{\mu\nu} \partial_{\nu} G(x;x') \Bigr) =
\delta^4(x-x') \quad \Longrightarrow \quad \theta(x) = \int d^4x' G(x;x')
S(x') . \label{putative}
\end{equation}
However, the retarded Green's function,
\begin{equation}
G^{\rm ret}(x;x') = \frac{H^2 \theta(\Delta t)}{4 \pi} \Bigl[ 2 \delta\Bigl(
y(x;x')\Bigr) + \theta\Bigl(-y(x;x')\Bigr) \Bigr] \; , \label{Gret}
\end{equation}
contains a $\theta$-function tail term which is nonzero throughout the
volume of the past light-cone. Because the source (\ref{source}) actually
{\it grows} as $t \rightarrow - \infty$, the integral (\ref{putative}), and
even its gradient, fail to converge.

Note that the electric field of (\ref{soln}) points from the $+q$ to the 
$-q$ charge and is nonzero throughout the full de Sitter manifold manifold,
\begin{equation}
F^{\chi 0} = \frac{q H^3}{4\pi} {\rm sech}^3(Ht) \csc^2(\chi) \; .
\end{equation}
This isn't at all what one gets by integrating the photon retarded Green's
function against the current density in a de Sitter invariant gauge,
\begin{equation}
A^{\rm ret}_{\mu}(x) = \int d^4x' \Bigl[{}_{\mu} G^{\rm ret}_{\nu}\Bigr](x;x') 
J^{\nu}(x') \; .
\end{equation}
One can recover the retarded Green's function from the Allen-Jacobson
propagator (\ref{AJprop}) by simply taking the imaginary part and multiplying 
by $-2 \theta(t\!-\! t')$,
\begin{eqnarray}
\lefteqn{\Bigl[{}_{\mu} G^{\rm ret}_{\nu}\Bigr](x;x') =  
\frac{H^2 \theta(\Delta t)}{4 \pi} \left\{ 2 \delta(y) - \frac{(8 \!-\! 
\frac43 y)}{(4 \!-\! y)^2} \theta(-y) \right\} \Bigl[{}_{\mu} g_{\nu}
\Bigr](x;x') } \nonumber \\
& & \hspace{2cm} + \frac{H^2 \theta(\Delta t)}{4 \pi} \left\{\frac{\frac43 
y}{(4 \!-\! y)^2} \theta(-y) \right\} \Bigl[{}_{\mu} n\Bigr](x;x') 
\Bigl[n_{\nu}\Bigr](x;x') . \qquad
\end{eqnarray}
The retarded Green's function is causal, so the response from it {\it 
vanishes} in the vast region of the full de Sitter manifold which is not
future-related to either of the source world lines. 

It turns out that the Allen-Jacobson Green's function {\it does} give the 
correct response within the open conformal coordinate patch, so a de Sitter 
invariant gauge can at least be imposed locally in electromagnetism. (I 
thank A. O. Barvinsky for correcting me about this, and I apologize to Allen 
and Jacobson for having said otherwise at the Deserfest.) The same does 
not seem to be true in gravity. Antoniadis and Mottola have shown that de 
Sitter invariant graviton propagators --- which are also much more complicated
than the one in my favorite gauge (\ref{GR}) --- lead to local violations 
of the linearized Einstein equations \cite{AM}! These violations are not
present when using my non-invariant propagator (\ref{gprop}) \cite{TW1}.

Note that the problem reconciling causality and the constraints is
classical. I advance for your consideration the folly of working much 
harder to quantize a formalism that doesn't even correctly reflect 
classical physics. When confronted with the causality obstacle de Sitter 
fanatics sometimes respond that the problem arises from the constraints 
not having been imposed throughout the initial value surface. When this 
is done the full system can be evolved just fine. I don't dispute this 
but it misses the point. The issue is not whether physics can be done on 
the full de Sitter manifold. There was never any doubt about that: 
(\ref{soln}) is the instantaneous Coulomb potential of Coulomb gauge.
The issue is rather whether or not physics can be done maintaining manifest 
de Sitter invariance. The answer is no. Imposing the constraints can always 
be subsumed into adding a surface gauge condition that breaks manifest de 
Sitter invariance.

Moving from classical to quantum field theory, recall that the condition for
getting enhanced quantum effects during inflation is massless particles
which are not classically conformally invariant. There are two such 
particles: the massless, minimally coupled scalar and the graviton.
Consideration of these particles is the {\it only} phenomonological 
justification for studying quantum field theory during inflation, so we
cannot dismiss them if they happen to violate aesthetic prejudices. As
it happens, the free quantum field theory of neither system possesses a 
normalizable, de Sitter invariant wave functional. This was proved long ago
for the massless, minimally coupled scalar by Allen and Folacci \cite{AF}. 
It can be seen for my graviton propagator (\ref{gprop}) by simply performing
a naive de Sitter transformation coupled with the compensating gauge
transformation needed to restore my noninvariant gauge (\ref{GR}) \cite{GK}.
Contrary assertions for gravitons are always based upon using de Sitter 
invariant gauges on the full manifold, which I have just shown to be 
incorrect.

The fact that the most interesting free quantum field theories have no
de Sitter invariant states means that the propagators of these fields
must break de Sitter invariance, not just through the gauge fixing
function but in a fundamental way. One can see this in the factor of 
$\ln(a a')$ on the second line of expression (\ref{DeltaA}) for $i
\Delta_A(x;x')$. There is no sense complicating a marginally tractable
formalism to respect a symmetry which is not there.

My final point is that even if manifestly de Sitter invariant propagators
had existed, the most interesting interactions would still break de 
Sitter invariance. I don't mean the interaction vertices would be 
noninvariant. They are manifestly invariant. What I mean instead is that 
higher order processes can involve integrals over interaction vertices. 
De Sitter invariant propagators would make the {\it integrands} invariant
but would not guarantee that the {\it integrals} were invariant. Consider 
the invariant volume of the past light-cone from some observation point 
$x^{\mu}$ back to the initial value surface (IVS) on which the prepared 
state was released,
\begin{equation}
V(x) \equiv \int_{IVS} \!\!\!\!\! d^4x' \, \sqrt{-g(x')} \, \theta(\Delta t) 
\, \theta\Bigl(-y(x;x') \Bigr) \; .
\end{equation}
The {\it integrand} is manifestly de Sitter invariant --- one inside
the past light-cone and zero outside --- but the {\it integral} grows as
the observation point is taken later and later after the state was 
released. One can see from the scalar retarded Green's function (\ref{Gret}) 
that this example is not artificial. Unsuppressed integrals over the volume 
of the past light-cone occur in many of these computations \cite{OW1}. They 
give factors of $\ln(a)$ every bit as important as the explicit ones from the 
de Sitter breaking terms of $i\Delta_A(x;x')$. It would not be far wrong to 
say that extracting these secular logarithms is the whole point of studying 
quantum field theory during inflation. 

\section{Conclusions}

Massless particles which are not conformally invariant can mediate 
interesting quantum effects during inflation. Even a single non-conformal 
massless particle can catalyze surprising processes which would otherwise 
not go \cite{RPW}. It is now possible to study this by modeling inflation 
as the open coordinate patch of de Sitter space, and by exploiting simple 
gauge fixing terms. 

This bothers de Sitter fanatics, who would prefer to work on the entire de 
Sitter manifold and to employ only de Sitter invariant gauge conditions.
That would not be easy because de Sitter invariant gauges complicate
propagators. It is also incorrect physically because a de Sitter invariant
gauge converts elliptic constraint equations into hyperbolic evolution 
equations. Whereas the former require a nonzero response throughout the 
manifold to a sufficiently distributed source, the later give zero response 
in the vast regions of de Sitter space which are not future-related to a 
source worldline. In some cases the use of a de Sitter invariant gauge even 
leads to violations of the original, gauge invariant field equations {\it
within} the region which is future-related to the a source worldline \cite{AM}!

The free quantum field theories of the two massless particles which are 
not conformally invariant admit no de Sitter invariant states. This means 
that the propagators of these fields cannot be de Sitter invariant 
\cite{AF,GK}. Even had all propagators been de Sitter invariant, interactions
would still break this invariance. The infrared logarithms which signal 
this breaking are at the heart of what makes quantum field theory during 
inflation potentially observable. I think we'd all rather have interesting 
quantum dynamics without symmetry than sterile dynamics with a beautiful 
symmetry. As Mick and Keith put it: ``You can't always get what you want. 
But if you try sometime you find, you get what you need!''

\section*{Acknowledgments}
It is a pleasure to acknowledge discussions and work on this problem with 
A. O. Barvinsky, T. Jacobson, E. O. Kahya, N. C. Tsamis and S. H. Yun. This 
work was supported by NSF Grant PHY-0244714 and by the Institute for 
Fundamental Theory at the University of Florida.

\end{document}